\input{psfig.sty}
\documentclass{aa}
\usepackage{amssymb}
\usepackage{txfonts}
\usepackage{natbib}

\newcommand\no{\nodata}

\def\ltsima{$\; \buildrel < \over \sim \;$}
\def\simlt{\lower.5ex\hbox{\ltsima}}
\def\gtsima{$\; \buildrel > \over \sim \;$}
\def\simgt{\lower.5ex\hbox{\gtsima}}
\begin{document}
\title{ Swift observations of GRB\,050904: the most distant cosmic
explosion ever observed.}

\author{G. Cusumano\inst{1}, V. Mangano \inst{1},
G. Chincarini\inst{2,3}, A. Panaitescu\inst{4},  D.N.
Burrows\inst{5}, V. La Parola\inst{1}, T. Sakamoto\inst{6,7}, S.
Campana\inst{2}, T. Mineo\inst{1}, G. Tagliaferri\inst{2}, L.
Angelini\inst{6}, S.D. Barthelmy\inst{6}, A.P. Beardmore\inst{8},
P.T. Boyd\inst{6}, L.R. Cominsky\inst{9}, C. Gronwall\inst{5}, E.E.
Fenimore\inst{4}, N. Gehrels\inst{6}, P. Giommi\inst{10}, M.
Goad\inst{8}, K. Hurley\inst{11}, S. Immler \inst{6}, J.A. Kennea\inst{5}, K.O.
Mason\inst{12}, F. Marshal\inst{6}, P. M\'{e}sz\'{a}ros\inst{5,13},
J.A. Nousek\inst{5}, J.P. Osborne\inst{8}, D.M. Palmer\inst{4},
P.W.A. Roming\inst{5}, A. Wells\inst{8}, N.E. White\inst{6}, B.
Zhang\inst{14} } 

\offprints{G. Cusumano: giancarlo.cusumano@ifc.inaf.it}

\institute{INAF-IASFPA, Via Ugo La Malfa 153, 90146
Palermo, Italy \and INAF -- Osservatorio Astronomico di Brera, Via
Bianchi 46, 23807 Merate Italy \and Universit\`a degli studi di
Milano-Bicocca, Dip. di Fisica, Piazza delle Scienze 3,  I-20126
Milan, Italy, \and Los Alamos National Laboratory, P.O. Box 1663,
Los Alamos, NM 87545, USA \and Department of Astronomy \&
Astrophysics, Pennsylvania State University,University Park, PA 16802
USA \and NASA/Goddard Space Flight Center, Greenbelt, MD 20771, USA \and
National Research Council, 2001 Constitution Avenue, NW, TJ2114,
Washington, DC 20418, USA \and Department of Physics and Astronomy,
University of Leicester, University Road, Leicester LE1 7RH, UK \and
Department of Physics and Astronomy, Sonoma State University,
Rohnert Park, CA 94928-3609, USA \and ASI Science Data Center, via
Galileo Galilei, 00044 Frascati, Italy \and UC Berkeley Space
Sciences Laboratory, Berkeley, CA 94720-7450 \and MSSL, University
College London, Holmbury St. Mary, Dorking, RH5 6NT Surrey, UK \and
Department of Physics, Pennsylvania State University, PA 16802, USA
\and Department of Physics, University of Nevada, Box 454002, Las
Vegas, NV 89154-4002, USA }


\date{Received: ..... ; accepted: ......}
\titlerunning{{\it Swift} observation of GRB\,050904}
\authorrunning{G. Cusumano et~al.}

\abstract
%
%
{
Swift discovered the high redshift (z=6.29) GRB\,050904
with the Burst Alert Telescope (BAT) and began observing with its narrow
field instruments 161 s after the burst onset.
This gamma-ray burst is the most distant cosmic explosion ever observed.
Because of its high redshift, the X-ray Telescope (XRT) and BAT simultaneous
observations provide 4 orders of magnitude of spectral coverage (0.2--150~keV;
1.4--1090~keV in the source rest frame) at a very early source-frame time
(22 s). The X-ray emission was monitored by the XRT up to 10 days after the
burst.
}
%
%
{
We present the analysis of BAT and XRT observations of GRB\,050904 and
a complete description of its high energy phenomenology.
}
%
%
{
We performed time resolved spectral analysis and light curve modeling.
}
%
%
{
GRB\,050904 was a long, multi-peaked, bright GRB
with strong variability during its entire evolution.
The light curve observed by the XRT is characterized by the presence of
a long flaring activity lasting up to 1-2 hours after the burst onset in
the burst rest frame, with no evidence of a smooth power-law decay
following the prompt emission as seen in other GRBs.
However, the BAT tail extrapolated to the XRT band joins the
XRT early light curve and the overall behavior resembles that
of a very long GRB prompt.
The spectral energy distribution softens with time,
with the photon index decreasing from $-$1.2 during the BAT observation
to $-$1.9 at the end of the XRT observation.
The dips of the late X-ray flares may be consistent with an underlying
X-ray emission arising from the forward shock and with the properties of
the optical afterglow reported by Tagliaferri et~al. (2005b).
}
%
%
{
We interpret the BAT and XRT data as a single continuous observation
of the prompt emission from a very long GRB.
The peculiarities observed in GRB\,050904 could be due to its origin
within one of the first star-forming regions in the Universe;
very low metallicities of the progenitor at these epochs may provide
an explanation.
}

\maketitle

\section{Introduction}
Gamma-Ray Bursts are bright flashes of high energy photons
that can last from about 10 milliseconds to 10 minutes. Their origin
and nature puzzled the scientific community for about 25 years
until 1997, when the first X-ray afterglow of long ($>$ 2 s
duration) bursts were detected \cite{costa97}, and the first optical
\cite{vanpara97} and radio \cite{frail97} counterparts were found.
These measurements established that long GRBs are typically at high
redshift (z$\sim1.6$) and are in sub-luminous star-forming host
galaxies \cite{bloom02}. They are likely produced in core-collapse
explosions of a class of massive stars that give rise to highly
relativistic jets (the collapsar model; \cite{mac01}). Internal
inhomogeneities in the velocity field of the relativistic expanding
flow lead to collisions between fast moving and slow moving fluid
shells and to the formation of internal shock waves \cite{rees94}.
These shocks are believed to produce the observed prompt emission in
the form of irregularly shaped and spaced pulses of gamma-rays, each
pulse corresponding to a distinct internal collision. The expansion
of the jet outward into the circumburst medium is believed to give
rise to ``external'' shocks, responsible for producing the smoothly
fading afterglow emission seen in the X-ray, optical and radio
bands \cite{meszaros97}.

The {\it Swift} \cite{gehrels04} X-ray Telescope (XRT; \cite{burrows05}) is
providing a growing number of unprecedented observations of the
early stages of GRB afterglow in the 0.2--10~keV X-ray band. The XRT
rapid ($\leq$ 2 min) response to the {\it Swift} Burst Alert Telescope
(BAT; \cite{barthelmy05a}) triggers has already led to the
discovery of rapid early X-ray declines followed by the smoother
``standard'' X-ray afterglow
components \cite{taglia05a, cusumano06a,barthelmy05b, vaughan06a}, dramatic flaring 
in the early
X-ray light curves of short \cite{fox05, barthelmy05c, campana06,
vaughan06b} and long bursts \cite{burrowscience, romano06,
falcone06,burrows05c,pagani06,morris06} and 
simultaneous peaks at the end of the
BAT observation (15--350~keV) and at the beginning of the
XRT observation (0.2--10~keV) of some long bursts (GRB\,050730, GRB\,050820a, 
GRB\,050822; see O'Brien et~al. 2006). 
Thanks to its fast response, and precise source localization, about 3$'$ in BAT, 3$''$--5$''$ in 
XRT and 0.3$''$ in UVOT (Roming et~al. 2006),
{\it Swift} is able to quickly alert ground-based telescopes to locate the
optical counterpart and get
redshift measurements before the object becomes too faint.

GRB\,050904 triggered the BAT on 2005 September 4 at
01:51:44 UT \cite{cumm05}. The burst was located on-board at RA$_{\rm
J2000}$=00h54m41s, Dec$_{\rm J2000}$=+14$^{\circ}$ 08$\arcmin$
17$''$ with an  uncertainty of 3$\arcmin$ radius
(90\% confidence level) and the spacecraft was quickly pointed towards it. 
The XRT on-board centroiding algorithm failed to detect the counterpart
due to the presence of a hot CCD detector column. The burst was long and bright  with 
duration $T_{90}$= (225$\pm$10) s and a 15--150~keV fluence of
(5.4$\pm$0.2) $\times$ 10$^{-6}$ erg cm$^{-2}$ \cite{taka05}.
UVOT did not detect the burst counterpart down to a 3 sigma upper limit
of about 21 mag in all its six filters \cite{cucchiara05}. 
The optical detection was first made by the robotic observatory TAROT
that begun to observe GRB050904 only 81 s after
the {\it Swift} trigger. A bright optical flare was detected during the 
prompt high-energy emission phase \cite{Boer06, Gendre06}.

Early photometry indicated a high redshift (z$>$5, \cite{reichart}).
A photometric redshift $z=6.1_{-0.12}^{+0.37}$ was measured by 
Tagliaferri et~al. (2005b) and confirmed by a Subaru
spectroscopic measurement of 6.29$\pm$ 0.01 \cite{kawai05}. A
break at $T_b$ = 2.6$\pm$1.0 days was also found in the J-band light
curve \cite{taglia05b}.

Such a high redshift  means that this explosion  happened 12.8 billion
years ago\footnote{We used standard cosmological parameters of
$\Omega_{\rm M}$=0.27, $\Omega_{\Lambda}$=0.73,
$H_0$=71\,km\,s$^{-1}$\,Mpc$^{-1}$}, corresponding to a time when the
Universe was  young ($\le 1 Gyr$), close to the re-ionization era \cite{beck01}.
This gave GRB\,050904 the distinction of being the most
distant cosmic explosion ever observed: the previous record for a
GRB was 4.5 \cite{andersen00}, the most distant quasar known is at a
redshift of 6.4 \cite{fan03}, and the most distant galaxy is at a redshift
of $\sim 7$ \cite{kneib04}.

Here we present the analysis of the BAT and XRT observations 
of GRB\,050904.  The details on data reduction
are in  Section 2; temporal and spectral analysis results are reported in sections 3 and
4, respectively. In section 5 we draw our conclusions.
Our results have already been shortly summarized by Cusumano et~al. (2006b), 
but this paper contains 
a complete description of the phenomenology of the GRB\,050904.
Compared to the Watson et~al. (2005) paper, that already presented 
the XRT data analysis, we add: {\it (i)} the detailed BAT data analysis,
{\it (ii)} complete and correct BAT and XRT light curves, 
{\it (iii)} time resolved spectral analysis on a finer temporal
grid, {\it (iv)} spectral analysis of BAT and XRT simultaneous data. 
Hereafter, errors are reported for a 90\% single parameter
confidence level.

\section{Observations and Data reduction}

The BAT data were analyzed using the standard BAT analysis
software included in the HEAsoft 6.0.4 package and described 
in the {\it Swift} BAT Ground Analysis Software 
Manual\footnote{http://swift.gsfc.nasa.gov/docs/swift/analysis/}.
BAT data from 306 to 525 s after the burst onset were telemetered  
in the masktag-lc observing mode that accumulates data in only four energy
bands. As a consequence, no spectral energy analysis was performed on
BAT data for this time interval.

\begin{figure*}[htb]
\label{fig1} \centerline{ \hbox{
\psfig{figure=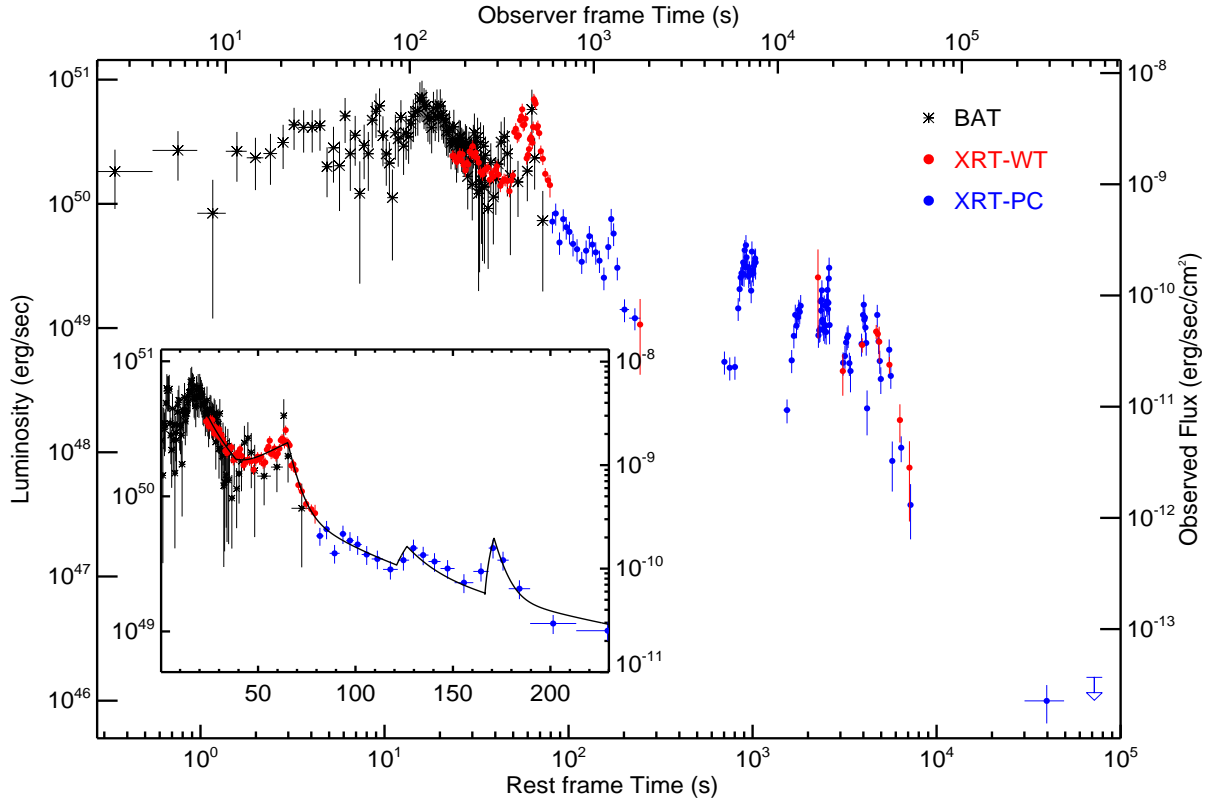,width=17cm,height=11cm,angle=-0,clip=} }}
\caption{The 0.2--10~keV light curve of GRB\,050904 as observed by the 
BAT and XRT.
The observed BAT count rates were extrapolated into the XRT 0.2--10~keV band using a conversion factor evaluated from the BAT best fit
spectral model (Table \ref{tab1}). The observed XRT count rates were
converted into  flux $F_{0.2-10~{\rm keV}}$ (scale on the right side) using the best fit 
spectral parameters listed in Table \ref{tab1}.
The equivalent isotropic luminosity (scale on the left side) was
calculated as
$F_{0.2-10~{\rm keV}}$ 4$\pi$ $d_L^2$ (1+z)$^{(\Gamma-2)}$
where $d_L$ is the luminosity distance and the last 
term is the K correction factor for emission modeled by a power law. 
The error bars are given by the quadrature sum of the count rate 
statistical error and the estimated uncertainties 
in the conversion factors. 
The inset shows BAT and XRT $F_{0.2-10~{\rm keV}}$ for the first orbit
and the best fit model for XRT data. 
Note that small differences in the 
the burst history  shown in the inset ($F_{0.2-10~{\rm keV}}$ units) and the main
picture, which shows the evolution in luminosity units, 
are due to the photon index dependence in the K correction factor and its
measurement in discrete time intervals.}
\label{fluxlc}
\end{figure*}
GRB\,050904 was observed by the XRT from 161 s up to 10 days after the
burst onset, overlapping the BAT observations for about 364 s.
The XRT observation started before the end of the high energy prompt
emission.
Data were accumulated in WT mode up to 573 s after the 
trigger time, while all the other
data were obtained in PC mode. In the WT mode only the central
$8\arcmin$ of the field of view is read out, providing one
dimensional imaging and full spectral capability with a time
resolution of 1.8 ms. The PC mode provides, instead, full spatial
and spectral resolution with a time resolution of 2.5 s.

XRT data were calibrated, filtered and screened using the XRTDAS package
included in the HEAsoft 6.0.4, as described in the
XRT Software User's Guide$^{2}$. 
Only observing time intervals with a CCD temperature below $-$47 
degrees Celsius were used. The total exposures after all the cleaning
procedures were 2.8~ks and 127.3~ks
for data accumulated in the WT and PC modes, respectively.
We used a 0--2 and 0--12 grade selection for data in
the WT and PC modes, respectively. Such a selection provides the best
combination of spectral resolution and detection efficiency.
The GRB was imaged far from the CCD hot columns and no
corrections for hot columns inside the photon extraction regions
was necessary.

The position of the burst was refined by on-ground analysis \cite{palmer05}.
The BAT burst position is RA$_{\rm J2000}$ = 00h54m53s, Dec$_{\rm
J2000}$ = +14$^{\circ}$$04\arcmin$52$^"$, with an uncertainty of
$2\farcm6$. This is $3\farcm9$ from the on-board position and 
$0\farcm54$ from the near--IR afterglow position \cite{Haislip05}. 
The XRT afterglow position derived with {\it xrtcentroid} (v0.2.7)
and including the latest boresight correction \cite{Moretti06}  
is RA$_{\rm J2000}$ = 00h54m50s.8, Dec$_{\rm J2000}$ =
+14$^{\circ}$05$\arcmin$09$\farcs$0, with an uncertainty of
$3\farcs5$. 
The XRT derived coordinates are $35$\farcs$9$ from the BAT ones
and $0$\farcs$4$ from the near--IR counterpart \cite{Haislip05}.

The BAT and XRT times are referred to the GRB\,050904 onset 
$T=$2005 Sep 4, 01:51:44.3 UT.

For the measured redshift $z=6.29$,
the 15--350~keV BAT band corresponds to a 109--2551~keV band in the burst rest frame
while the 0.2--10~keV XRT band corresponds to a 1.4--73~keV band. 
The observed timescales are stretched by a factor $(1+z)$ with respect 
to the rest frame ones.
In the following the GRB phenomenology is presented and discussed
from the point of view of the source rest frame.


\section{Timing Analysis}

The timing analysis for the XRT data was performed by selecting
events from a region centered on the source with a radius of 
20 (47.2) and 35 (82.6) pixels (arcsec) for WT and PC data, 
respectively. 
The background was estimated from regions
sufficiently offset ($>$ 2 arcmin) from the source position to avoid
contamination from the PSF wings and free from contamination by other
sources.

The intensity of the source caused pile-up in the PC data
up to 8~ks from the burst onset. 
The pile-up correction was performed by excluding 
a central region of 4 pixels radius and dividing the
extracted count rate by the fraction of lost point spread function (PSF; 58\%).
For the rest of the observation we used the full circular 
extraction region.

WT data were also corrected for the fraction of PSF not included in the
extraction region (7\%).

Figure \ref{fluxlc} shows the evolution of the GRB flux and luminosity.
The BAT light curve displays three main peaks: two short peaks ($\sim$2 s long) 
at T+3.8 and at T+9.3 s, and a main
long-lasting peak at $\sim$T+13.7 s, where T is the time of
the burst onset. Emission in the BAT energy range continues up to
almost T+77.7 s with a weak peak at $\sim$T+65 s,
coincident with the first peak of the XRT light curve. 
The BAT and XRT light curves overlap 
between T+23 and T+69 s. The early XRT light curve 
shows a steep decay with a slope $\alpha=-2.07\pm 0.03$ 
with three flares superimposed at T+65 s, T+126 s and T+171
s. These flares can be modeled by a linear rise 
lasting 26.6, 5.3 and 4.7 s, 
plus an exponential decay with decay time of 4.5, 10.98 and 5.2 s, 
respectively. 
The best fit model of the first orbit of the XRT light curve
is shown in the inset of Fig. \ref{fluxlc}. 
Although interrupted by observing constraints imposed by
the {\it Swift} orbit, the light curve from GRB\,050904 reveals highly
irregular rate variations likely due to the presence of flares up to
T+1.5 hours. At later times the flaring activity is not detected and
only a residual emission, 10$^{5}$ times lower than the initial
intensity, is visible.
Note that the XRT light
curve presented in Watson et~al. (2005) shows an evident discontinuity
at the end of the WT observation segment likely caused by an error in his 
flux conversions for the following PC mode data, all of which are 
systematically too low by about a factor of four.

\begin{figure}[htb]
\label{fig2} \centerline{ \hbox{
\psfig{figure=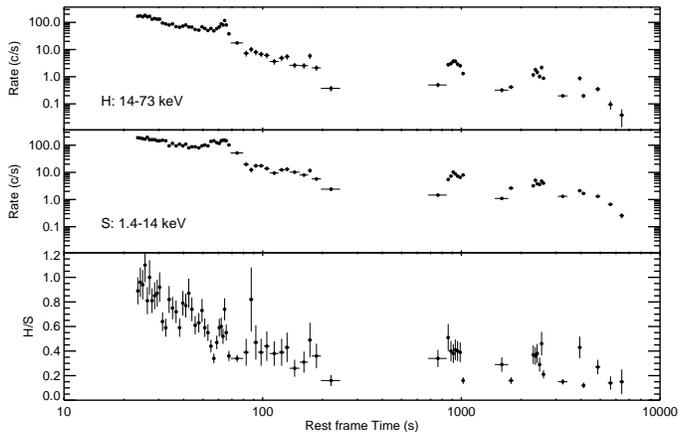,width=9.9cm,angle=-0,clip=} }}
\caption{Hardness ratio evolution of GRB\,050904. The top and middle
panels show the count rate evolution of XRT data in two different
energy bands in the rest frame. 
}
\label{hrlc}
\end{figure}

Figure \ref{hrlc} shows the rest frame XRT light curves of GRB\,050904 in the
1.4--14~keV and 14--73~keV bands (top and middle panels). 
The two light curves are binned before conversion to the rest frame 
in order to have at least 40 counts per bin in both bands. 
The hardness ratio (bottom panel) is defined as H/S, 
where H (hard) and S (soft) are the high energy and
the low energy bands.
The H/S plot reveals a significant shift to softer energies 
with time during the first 80 s, 
with the exception of the flaring episode at T+65 s,
where the hardness ratio peaks, too. Later emission shows no
evidence of further softening until the tail of the last flare
around T+5500 s.

\section{Spectral Analysis}
\begin{figure}[htb]
\label{fig3} \centerline{ \hbox{
\psfig{figure=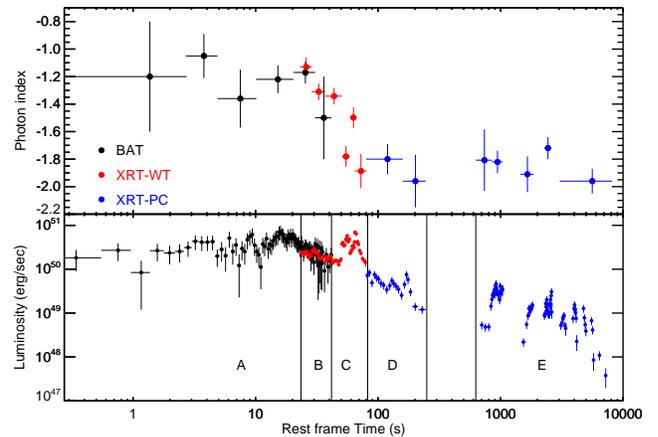,width=9.9cm,angle=-0,clip=} }}
\caption{Spectral evolution of GRB\,050904. The top panel
illustrates how the photon index $\Gamma$ of GRB\,050904 changes
during the observation. In the bottom panel the burst evolution is
plotted to show how the time intervals for spectral analysis were
selected. Vertical bars indicate the time intervals selected for BAT and XRT spectral analysis
whose results are reported in Table 2 and Fig. 4.}
\label{alphalc}
\end{figure}

The spectral analysis of GRB\,050904 was performed by selecting two
sets of time intervals for the BAT and XRT observations,
corresponding to characteristic phases of the light curve evolution.
The BAT spectra were accumulated in the 14--150~keV observed band in
six time intervals up to 41.6 s from the burst onset (Table 1).
No BAT spectrum was accumulated from T+41.6 to T+72 s because BAT
data were in the masktag-lc observing mode with only four energy bands.
The XRT spectra were accumulated in thirteen time intervals 
from T+23 s to T+8173 s from the burst onset (Table 1).
The WT and PC spectra were extracted from the same regions used for the
timing analysis.
Instrumental energy channels below 0.3~keV and above 10~keV for PC and WT
spectra were ignored and the background was evaluated in
regions sufficiently offset ($>$ 2 arcmin) from the source position to
avoid contamination from the PSF wings and free of contamination from other 
sources in the field of view. Moreover, energy channels between 0.5 and
0.6 keV were excluded because of the presence of a sistematics in 
such a range due to a time dependent energy scale 
problem at low
energy\footnote{http://heasarc.gsfc.nasa.gov/docs/swift/analysis/}; 
a sistematics in the gain offset produces a negative residual in the 0.5
keV region increasing in the fit procedure the N$_{\rm H}$ estimation.
The BAT spectra were modeled with a power law
with photon index $\Gamma$ ($F(E) \propto E^{\Gamma+1}$)
while the XRT spectra were modeled with a power law plus
two absorption components:
one for the intrinsic absorption in the host galaxy and one for the
Galactic absorption. The latter was fixed to the line-of-sight value
of 4.93$\times10^{20}$ cm$^{-2}$ \cite{Dickey90}. 
The model gave acceptable $\chi^2_\nu$ values for all the selected intervals (Table 1).
More complex models, such as a Band function \cite{band93},
cannot be constrained by the data.
Figure~\ref{alphalc} shows the evolution with time of 
the photon index $\Gamma$. 
The BAT
spectra have $\Gamma\sim-1.2$, consistent with typical values
of the $\alpha_{\rm Band}$ parameter of the Band model \cite{pree00}.
This strongly suggests that the BAT observes the low energy part of
the Band function and that the peak energy of the GRB spectrum is
above $150 \times (1+z)$~keV in the source rest frame. If we exclude
the spectrum of the first XRT flare at T+65 s, the XRT photon
indices show a clear decreasing trend from about $-1.2$ to about
$-1.9$ in the first T+200 s. No further spectral evolution is
present in later XRT data, in agreement with the hardness ratio
curve. The BAT and XRT photon indices are in good agreement in the
overlapping region.
We also detected in the XRT WT spectra (T $<$ T+67.1 s), a highly 
significant absorption in excess of the Galactic
value. The intrinsic absorption column decreases with time at high
significant level.
In the time intervals where the burst was observed in PC mode
(T $>$ T+67.1 s), due to the decreased statistics, only upper
limits were measured. 
Table \ref{tab1} shows the best fit parameters for each of the selected time
intervals.

\begin{figure}[htb]
\label{fig4} \centerline{ \hbox{
\psfig{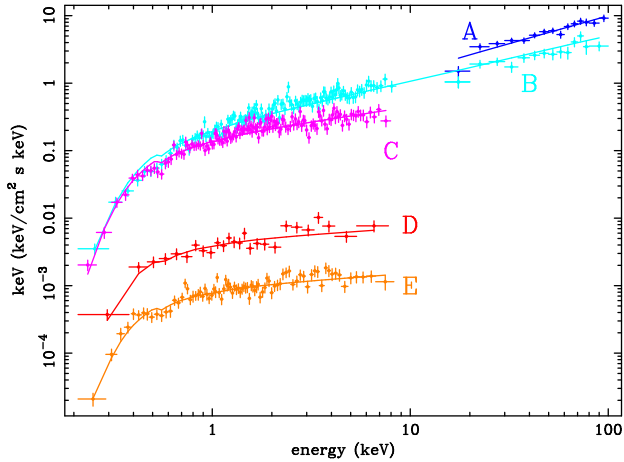} }}
\caption{Spectral evolution of GRB\,050904; BAT and XRT $E^2$F(E)
spectra (equivalent to $\nu$F($\nu$) spectra)
deconvolved from the detector response for five selected time intervals (see bottom
panel in Fig. 3). 
}
\label{EEF}
\end{figure}

The GRB\,050904 spectral evolution  is also quite
evident in Fig.~\ref{EEF} where the $E^2$F(E) deconvolved 
spectra (equivalent to $\nu$F($\nu$) spectra)
selected in 5 ad hoc intervals (bottom panel in Fig. 3)
are shown together with the best fit spectral models.
The first spectrum (A) is accumulated in the time interval T$-$1.43 s $-$
T+23.2 s, when only BAT observed the burst. The second spectrum (B) is
from T+23.2 s to T+41.7 s when the GRB is simultaneously
observed
by BAT and the XRT. The following spectra (C, D,
E) refer to the time intervals T+41.6 s $-$ T+82.3 s, T+82.3 s $-$
T+224
s and T+628 s $-$ T+8163, respectively, when only the XRT detected the
burst. 
The energy distribution is
clearly softening with time starting from interval C.
The fit in the broad energy band of interval B is also well modeled by an
absorbed power law while a Band function \cite{band93},
cannot be constrained by the data.
Table \ref{tab2} shows the best fit parameters for each of the selected time
intervals.
\begin{table*}[ht]
\begin{center}
\small{
\begin{tabular}{rrrccccc}
\hline Interval&\multicolumn{2}{c}{Time (s)} & $N_H$ ($10^{22}$ $cm^{-2})$ & $\Gamma$  & $\chi^2_\nu$
(dof)
        &\multicolumn{2}{c}{Flux ($10^{-9}$erg cm$^{-2}$ s$^{-1}$)} \\
  & Start & Stop & &
& 0.2--10 keV & 15--350 keV \\
\hline
BAT 1&$-$1.43  & 2.69& -- &$-$1.2 $\pm$0.4  &1.2 (57)  & 1.4&22.9\\
2    & 2.69  & 4.89& -- &$-$1.05$\pm$0.16 &0.86 (57) &3.2&90.8\\
3    & 4.89  & 10.1& -- &$-$1.36$\pm$0.21 &0.97 (57) &3.4& 30.9\\
4    & 10.1  & 20.4& -- &$-$1.17$\pm$0.08 &0.95 (57) &3.6& 66.8\\
5    & 20.4  & 30.6& -- &$-$1.22$\pm$0.10 &0.93 (57) &3.0& 45.7\\
6    & 30.6  & 41.6& -- &$-$1.5 $\pm$0.3  &0.88 (57) &2.0& 9.9\\
\hline
XRT 1&23.2  & 28.7& 5.73$\pm$4.2&$-$1.19$\pm$0.1  & 0.77 (62) &3.5& --  \\
2    &28.7  & 36.9& 5.5$\pm$2.5 &$-$1.34$\pm$0.08 & 0.98 (95) &2.5& --  \\
3    &36.9  & 50.6& 3.4$\pm$2.2 &$-$1.33$\pm$0.08 & 0.78 (89) & 1.3& --  \\
4    &50.6  & 58.8& 7.7 $\pm$4.5 &$-$1.85$\pm$0.1  & 1.12 (56) & 1.4& --  \\
5    &58.8  & 67.1& 4.2$\pm$2.0 &$-$1.50$\pm$0.09 & 1.14 (73) & 1.7& --  \\
6    &67.1  & 79.8& 1.5$\pm$1.4 &$-$1.86$\pm$0.13 & 0.94 (37) & 0.54 &--  \\
7    &79.8  &159.4& $<$6.4 &$-$1.80$\pm$0.15 & 1.12 (23) & 0.12& --  \\
8    &159.4 &244.4& $<$6.4 &$-$1.97$\pm$0.24 & 0.90 (7)  & 0.05& --  \\
9    &628   & 848 & $<$5.2 &$-$1.80$\pm$0.24 & 0.92 (7)  & 0.02& --  \\
10   &848   &1040 & $<$6.8 &$-$1.86$\pm$0.14 & 0.90 (35) & 0.08& --  \\
11 &1452 &1863    & $<$5.8    &$-$2.01$\pm$0.22 & 0.80 (17) & 0.02& -- \\
12 &2275 &2618    & $<$6.9 &$-$1.90$\pm$0.14 & 1.26 (47) & 0.04& -- \\
13 &3045 &8173    & $<$4.0    &$-$1.97$\pm$0.12 & 1.24 (35) & 0.008& -- \\
\hline
\end{tabular}}
\end{center}
\caption{BAT and XRT spectral analysis results. The BAT fluxes in the XRT 
band are extrapolated from the best fit models. Quoted errors are at the 90\%
confidence level.}
\label{tab1}
\end{table*}

We also evaluated the contribution to the total fluence in the 1.4$-$73\,keV  
band of the three flares (T+65, T+126 and T+171 s) superimposed on the early 
XRT light curve. The fluence values over the continuum are
$(1.2 \pm 0.08) \times 10^{-6}$, $(4.7 \pm 0.5) \times 10^{-8}$ and 
$(5.8 \pm 0.6) \times 10^{-8}$\,erg\,cm$^{-2}$, respectively. 
The fluence of the XRT continuum over the first orbit (i.e. from 23.2 to 244.4\,s) 
is $(4.9 \pm 0.3) \times 10^{-6}$\,erg\,cm$^{-2}$. 
The extrapolated 1.4--73\,keV BAT fluence in the time interval from the burst onset
to the start of the XRT observation is $(4.1\pm 0.2)\times10^{-6}$\,erg\,cm$^{-2}$.
The three XRT flares are therefore 5\%, 1\% and 1\% of the total 1.4--73\,keV
emission observed up to T+244 s.
The 1.4--73\,keV fluence in the remaining part of the XRT observation is 
$1.8$ $\times$ $10^{-6}$\,erg\,cm$^{-2}$. This value is only a lower limit 
because of the observing gaps.

\begin{table*}[ht]
\begin{center}
\small{
\begin{tabular}{crrrccccc}
\hline 
& Interval&\multicolumn{2}{c}{Time (s)} &$N_H$ ($10^{22}$ $cm^{-2})$ & $\Gamma$  & $\chi^2_\nu$ (dof) &\multicolumn{2}{c}{Flux ($10^{-9}$erg cm$^{-2}$ s$^{-1}$)} \\
&  & Start & Stop & & & & 0.2--10 keV & 15--350 keV \\
\hline
BAT &A    & $-$1.43  & 23.2 & -- & $-$1.19 $\pm$0.07  & 0.96 (57)  & 3.1 & 52.8\\
BAT+XRT &B& 23.2     & 41.6 & 7.5$\pm$1.5& $-$1.37$\pm$0.02 &1.0 (189)& 2.6 &11.9\\
XRT &C    & 41.6     & 82.3 & 4.7$\pm$1.2& $-$1.63$\pm$0.05 &1.19 (205) &1.33& --\\
XRT &D    & 82.3     & 224  & $<$ 2.1 & $-$1.7$\pm$0.1 & 1.24 (22) &0.03 & --\\
XRT &E    & 628      & 8163 & $<$ 2.9 & $-$1.74$\pm$0.07 & 1.243 (80) &0.006& --\\
\hline
\end{tabular}}
\end{center}
\caption{
BAT and XRT spectral analysis results of relative to 5 ad hoc selected intervals of the GRB
evolution (see bottom panel in Fig. 3). The BAT fluxes in the XRT
band are extrapolated from the best fit models. Quoted errors and upper limits are at the 90\%
confidence level.
}
\label{tab2}
\end{table*}

\section{Discussion}

GRB\,050904 was a long, multi-peaked, bright GRB with strong 
X-ray flaring activity lasting up to 1--2 hours in the source rest frame
(or 5$--$6~$\times 10^4$ s in the observer frame).
X-ray flares are detected in nearly half of the {\it Swift} long 
bursts (e.g. GRB\,050406 \cite{burrowscience,romano06}, GRB\,050502
\cite{burrowscience,falcone06}, 
GRB\,050607 \cite{pagani06}, GRB\,050713A \cite{morris06}). 
The rise time and decay time of these flares seen
at lower z are frequently  very fast with a ratio between the 
duration and peak time $\delta t/t \ll 1$. These features make
difficult to explain these flares with mechanisms associated 
with the external shock (Zhang et~al. 2006)

The variability of GRB\,050904 X-ray light curve is even more dramatic than the 
typical {\it Swift} afterglow showing flares, and 
the amplitude and rise/fall times of the flares are consistent
with the behavior of nearby (z $\leq$1) long GRBs (Fishman \& Meegan
1995). 
This suggests to interpret them as late
internal shocks related to central engine activity. 
In this scenario they would have the same origin as the prompt 
gamma-ray emission \cite{zhang06,nousek06,obrien06}. 
This would require that the
central engine remain active up to at least 5000 s,
consistent with the collapsar model \cite{mac01}, which allows central
engine activity for up to a few hours. 
Then, BAT and XRT have likely recorded a single continuous
observation of long lasting prompt emission where the standard
X-ray afterglow component is hardly detectable because its contribution
is drowned by the intense and long lasting activity of the central
engine.

TAROT observed a flare in the optical band \cite{Boer06,
Gendre06} which is simultaneous with the X-ray flare at T+65 s. Later
( $>$ T+1200 s) optical and infrared observations \cite{taglia05b} are 
too sparse to test if the flaring activity observed at later times
in X-rays is indeed present. The multiwavelength fit by
Tagliaferri et~al. (2005b) suggests that the optical data taken after
T+1200 s are in agreement with standard afterglow emission and jet
lateral spreading  at about T+80 ks. 
Physical interpretations to explain the simultaneous X-ray and optical flare 
at T+65 s are extensively discussed in Gendre et~al. (2006). They showed
that delayed external shock from a thick shell (Piro et~al. 2005), synchrotron
emission from reverse shock (Fan \& Wei 2005),
inverse Compton emission from reverse shock (Kobayashi et~al. 2005) and
inverse Compton from late internal shock (Wei et~al. 2006) cannot 
satisfactory explain the simultaneous X-ray and optical flare. 
On the other hand, the late internal shock model proposed by Zou et~al. (2006) 
could account for the broad band data of the first flare. 

In the time interval from T+23 to $T+244$ s, the observed
intensity underlying the XRT flares decays as $t^{\alpha}$ with
$\alpha$ = $-$2.07$\pm$0.03.  An initial steep decay of the X-ray emission has
been observed in many other GRBs detected by {\it Swift}
\cite{taglia05a,nousek06,obrien06}. The decay slope 
together with the XRT energy index $\beta = \Gamma+1 \sim -0.2$  
measured up T+50 s are in good
agreement with the interpretation of the observed emission as due to
high-latitude emission \cite{kum00}. 
In this model, the tail of a peak is the emission
from the shocked gas moving at angle $\theta$ $>$ 1/$\gamma$ (where
$\gamma$ is the Lorentz factor) relative to the observer.
The higher the angle $\theta$, the later the photon arrival-time
and the weaker the relativistic beaming of the emission,
leading to a $t^{\alpha}$ decay with $\alpha=\beta-2$.
After T+50 s, due to the decrease of $\beta$ to about $-0.7$,
the predicted slope would be steeper than the measured $-2$. This
deviation could be reconciled with the high-latitude emission assuming
that the delayed radiation from the outer parts of the emitting
curved shell is softer and brighter than the radiation along our 
line of sight \cite{kum06}.
 
The decrease of the photon index around T+50 s could be
interpreted as an indication of a shift of the spectral peak energy
($E_p$) towards lower
energies, but poor statistics and the narrowness of the XRT energy
range do not allow us to verify this hypothesis.

We detected highly significant absorption in excess of the Galactic
value. The host column density decreases with time at a high
significance level. This is consistent with the idea that the
circum-burst absorbing material is photoionized by the high-energy
photons of the burst (Perna \& Loeb 1998). Evidence for such a decrease has been found
for GRB\,980329 \cite{frontera00} and GRB000528 \cite{frontera00} but 
neither of them has a comparable data quality. 
We do not confirm evidence found by Watson
et~al. (2006) of an increase of the column
density at the peak of the first X-ray flare.

Our lack of knowledge concerning the peak energy of the BAT and XRT
spectra does not allow a precise estimate of the total energy
released by GRB\,050904. However, we  can calculate lower and upper
limits to the isotropic-equivalent radiated energy $E_{\rm iso}$ up to
244 s from the burst onset, i.e. including contributions from
the first three XRT flares. To evaluate the lower limit to $E_{\rm iso}$
we integrated the best fit power law spectral energy distributions
in the $(1-200)\times$(1+z)~keV band and in the $(1-10)\times$(1+z)~keV band for the BAT and XRT, respectively. The standard
energy range $1-10^4$~keV (rest frame) was used to evaluate the upper
limit to $E_{\rm iso}$. 
We obtained $6.6\times 10^{53}$ erg $
< E_{\rm iso} < 3.2\times 10^{54}$ erg. Additional contributions from
the later flare portions are only a few percent in both the upper and
lower limit. The large 
isotropic equivalent energy of this burst is in agreement
with the Amati relation \cite{amati02} with an $E_p$ of about 1500~keV in the rest frame. This is consistent with our non-detection of
a peak energy in BAT spectral fit.

The break observed in the optical and infrared afterglows
\cite{taglia05b} 
at $T_b$ = 2.6$\pm$1.0 days (observer frame) and the range of
$E_{\rm iso}$ evaluated above imply a jet half-opening angle 
$\vartheta_{\rm jet}$ between $2\degr$ and  $4\degr$, assuming a radiative 
efficiency $\eta$ = 20\% and
a circumburst medium density n = 3 cm$^{-3}$. This angle estimate
 is consistent
with those obtained by modelling optical light-curve breaks observed in 10 pre-{\it Swift}
GRBs (Panaitescu 2005).
It corresponds to a collimation-corrected energy $E_{\gamma}$ between 
0.4 and 4.4$\times 10^{51}$ erg.  This is well within the $E_{\gamma}$ distribution of
GRBs with known redshift \cite{frail01, bloom03}.
Consistency with the Ghirlanda relation \cite{ghir04}
constrains the rest frame peak energy of the average spectrum
to be between 560 and 1300~keV.

In the optical afterglow, the measured pre-break slope decay 
$\alpha_o$ =  $-$0.7 $\pm$ 0.2 (Tagliaferri et~al. 2005) requires an
electron index of p = (4/3)$\alpha_o$+1 $\sim$ 1.9$\pm$0.3 in a local uniform
interstellar medium. The expected intrinsic optical spectral slope should be
$\beta_o$ = (p$-$1)/2 $\sim$ $-$0.45$\pm$0.15 while the observed one is
$\beta_o$ = $-$1.25$\pm$0.15 (Tagliaferri et~al. 2005). This discrepancy
could be naturally explained assuming the presence of a bit of dust 
($A_V$ = (1.25$-$0.45)/(1+z) = 0.1) in the local host medium.
In the X-ray regime the slowest decay of a possible
underlying continuum, inferred by fitting the lowest points in Fig. 1
after T+1000 s, 
is $\alpha_x$ $\sim$ $-$1.2. This is marginally consistent with the slope 
expected for the forward shock emission for the value of electron index
inferred from optical data: $\alpha_x$ = (3p$-$2)/4 $\sim$ 0.95$\pm$0.20,
under the hypothesis that the  cooling frequency is between optical and X-rays.
The expected spectral index in X-rays is $\beta_x$ = p/2 $\sim$ $-$1.0$\pm$0.1, 
consistent with the observed $\beta_x$= $-$ 1.0$\pm$0.2 after T+1000s.
The continuum component inferred by the fitting the the lowest points in
Fig. 1 is, therefore, marginally consistent with a standard forward
shock emission and the properties of the optical emission reported 
by Tagliaferri et~al. (2005). However, we cannot prove that the
estimated underlying X-ray continuum and the late optical light curve 
are the same forward shock emission because the observational gaps 
do not allow to see the minima of the X-ray light curve.

Figure 5 shows how GRB\,050904 would appear in the {\it Swift} 0.2--10~keV band
 if it were at redshifts other than z=6.29. 
The observed intensity varies inversely with the square of the  
(1+z) factor. 
A further dependence on z is due to the  
K correction that accounts for the redshift dependence of the
luminosity in a given wavelength band. The observed timescale
undergoes different stretching factors with respect to the rest frame:
the burst would appear longer at higher redshift.
The horizontal line in Fig. 5  gives an indication of the sensitivity limit 
of BAT, while the vertical line  marks the start of the XRT follow-up. 
Starting from a redshift lower than $\sim$ 2, BAT would
have observed all later flares. GRB\,050904 would have 
been detected up to a distance corresponding to z$\sim$10.
Comparing the intensity of other GRBs observed by {\it Swift} (see Fig.3 in
O'Brien et~al. 2006) with the flux that we would observe if
GRB\,050904 had exploded at a redshift lower than 2 ( $\sim$ the average
redshift of the {\it Swift} GRBs) we see that this burst was intrinsically bright 
in X-rays. 

After one year of operations, {\it Swift} has detected four confirmed
high-redshift (\simgt 4) GRBs (out of 24 with known redshifts).
Figure 6 shows the K-corrected 0.2--10~keV luminosity  evolution of
these GRBs, including both BAT and XRT data set. BAT light curves are
obtained by extrapolation of the $15-150$~keV light curve to the 
XRT energy band.
In addition to GRB\,050904 this sample includes GRB\,050730 
at z=3.969 (Chen et~al. 2005), GRB\,050505 at z=4.27 (Berger et~al. 2006) 
and GRB\,050814 at z=5.3 (Jakobsson et~al. 2006).
All of the bursts are exceptionally
luminous and long-lasting (as measured in the source rest-frame) and
are among the brightest GRBs ever
observed. Their exceptional intensity is not due to selection effects,
since the $15-150$~keV burst fluxes are well above the BAT detection threshold.  Rather,
their unique properties could be due to their likely origin within some
of the first star-forming regions in the Universe; Woosley \& Heger
(2006) suggested that the very low metallicities of the progenitors at
these epochs may provide an explanation. 
A more reliable conclusion about  systematic differences and similarities in 
luminosities and durations of high redshift
GRBs will require an increase in the sample size, which should come in future years of
{\it Swift} operation.  

Detecting high-redshift GRBs with {\it Swift}, and measuring their
redshifts with ground-based spectroscopy, is of substantial interest
because of the link between long-duration GRBs and the star formation
rate. The GRBs with measured redshift can be used to infer the cosmological 
star formation
history, with relatively minor (or in any case unique) selection effects
by comparison to other methods (Porciani \& Madau 2001; Lamb \& Reichart
2000; Bromm \& Loeb 2002; Natarajan et~al. 2005).  A preliminary estimate
of the star formation rate derived from {\it Swift} bursts (Price et~al. 2006)
shows, within current broad uncertainties, a flat or (at the highest
redshifts) slowly-declining star formation rate, consistent with
results obtained from color-selected galaxy observations (Bunker et~al. 2004). 

\begin{figure}[htb]
\label{fig5} \centerline{ \hbox{
\psfig{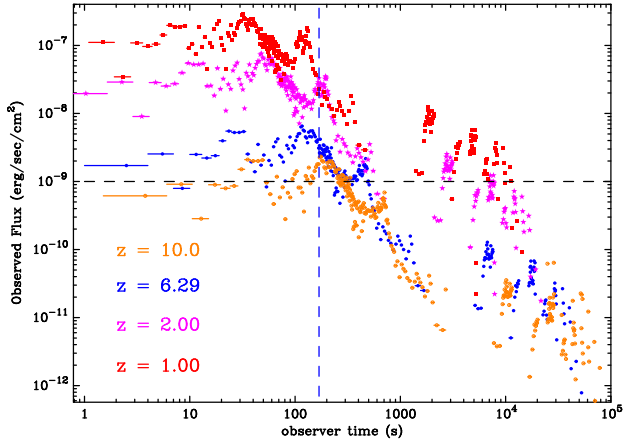} }}
\caption{GRB\,050904 light curves in the $0.2-10$~keV band as they would be observed if
the burst were located  
at different redshifts. Each light curve includes both BAT and XRT
extrapolations. The horizontal line gives an indication of the sensitivity limit
of BAT, while the vertical line  marks the start of the XRT follow-up in
the observing frame.
}
\label{zvarying}
\end{figure}

\begin{figure}[htb]
\label{fig6} \centerline{ \hbox{
\psfig{figure=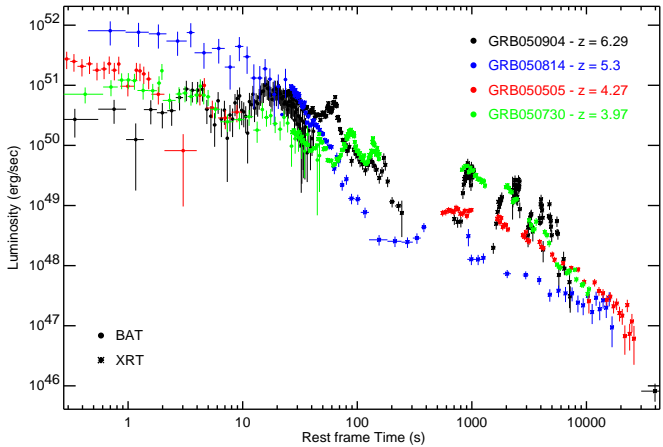,width=9.9cm,angle=-0,clip=} }}
\caption{The K-corrected 0.2--10~keV time histories of the high-redshift (\simgt 4) GRBs
observed by {\it Swift}. Each light curve includes both XRT and BAT data
extrapolated to the XRT energy range. The K correction was performed 
for an average photon index for simplicity.
}
\label{highz}
\end{figure}

\section{Summary and conclusion}

We have presented the results of the analysis of BAT and XRT 
observations of  the high redshift (z=6.29) GRB\,050904.
The GRB light curve is characterized by the presence of
a long flaring activity lasting up to 1-2 hours after the burst onset in
the burst rest frame, with no evidence of a smooth power-law decay
following the prompt emission as seen in other GRBs.
We interpret the overall phenomenology of GRB\,050904 as long lasting
prompt emission where the standard
X-ray afterglow component is hardly detectable because its contribution
is drowned by the intense and long lasting activity of the central
engine.
The spectral energy distribution softens with time,
with the photon index decreasing from $-$1.2 during the BAT
observation to $-$1.9 at the end of the XRT observation.
The dips of the late X-ray flares may be consistent with an underlying
X-ray emission arising from the forward shock and with the properties of
the optical afterglow reported by Tagliaferri et~al. (2005b).

Highly significant absorption in excess of the Galactic value has been
detected. The intrinsic hydrogen-equivalent colunm density shows a significant
decreasing with time that we interpret as due to the photoionization of the 
circum-burst absorbing material by the high-energy photons of the burst 
causing a graduale reduction of the opacity.

We have calculated lower and upper limits to the isotropic-equivalent
radiated energy $E_{iso}$ up to 244 s from the burst onset, i.e.
including contributions from the first three XRT flares.
We obtained $6.6\times 10^{53}$ erg $ < E_{iso} < 3.2\times 10^{54}$
erg. This range of $E_{iso}$ and the break observed in the optical and
infrared afterglows imply a jet half-opening angle
$\vartheta_{\rm jet}$ between $2\degr$ and  $4\degr$, assuming a radiative
 efficiency $\eta$ = 20\% and a circumburst medium density n = 3 cm$^{-3}$.

\begin{acknowledgements}
The authors acknowledge support from ASI, NASA and PPARC.
KH is grateful for support under NASA grant FDNAG5-9210.
We also acknowledge the anonymous referee for very useful suggestions.

\end{acknowledgements}

\end{document}